\documentclass[10pt,a4paper,oneside,openbib]{article}

\usepackage{epsfig}
\usepackage{fullpage}
\usepackage[T1]{fontenc}
\usepackage[latin1]{inputenc}
\usepackage{bbm}
\usepackage{amsgen}[1995/01/01]
\usepackage{amsfonts}
\usepackage{amssymb}
\usepackage{amsbsy}

\def\rmi{\mathbbm{i}}
\def\rme{{\rm e}}
\def\rmd{{\rm d}}
\def\rmi{{\rm i}}

\def\atan{{\rm atan\, }}

\renewcommand{\title}[1]{\topsep=0pt\begin{flushleft}\LARGE\bf#1\end{flushleft}\vspace{12pt}}
\renewcommand{\author}[1]{\topsep=0pt\begin{flushleft}\large\rm#1\end{flushleft}}
\newcommand{\address}[1]{\topsep=0pt\begin{flushleft}\footnotesize\it#1\end{flushleft}\vspace{12pt}}

\renewcommand{\date}[1]{\topsep=0pt\begin{center}#1\end{center}\vspace{12pt}}

\begin{document} 

\title{Phase dynamics and particle production in  preheating}

\author{T. Charters${}^{\dag}$, A. Nunes${}^{\ddag}$, J. P. Mimoso${}^{\S}$}

\address{${}^{\dag}{}$
Departamento de Mec\^anica/Sec\c c\~ao de Matemática\\
Instituto Superior de Engenharia de Lisboa\\ Rua Conselheiro 
Em\'{\i}dio Navarro, 1, P-1949-014 Lisbon, Portugal\\
Centro de F\'{\i}sica Te\'{o}rica e Computacional
da Universidade de Lisboa \\
Avenida Professor Gama Pinto 2, P-1649-003 Lisbon, Portugal}

\address{${}^{\S}$${}^{\ddag}$
Departamento de F\'{\i}sica, Faculdade de Ci\^encias da Universidade de Lisboa \\
Centro de F\'{\i}sica Te\'{o}rica e Computacional
da Universidade de Lisboa \\
Avenida Professor Gama Pinto 2, P-1649-003 Lisbon, Portugal
}

\address{${}^\dag$tca@cii.fc.ul.pt, ${}^\ddag$anunes@ptmat.fc.ul.pt, ${}^\S$jpmimoso@cii.fc.ul.pt}

{\footnotesize PACS: 98.80.Cq}

{\footnotesize \today}

\begin{abstract}
We study a simple model of
a massive inflaton field $\phi$ coupled to another scalar filed $\chi$
with interaction term $g^2\phi^2\chi^2$.
We use the theory developed by Kofman {\em et al.} \cite{Kofman:1997b}
for the first 
stage of preheating to give a full description of the dynamics of the $\chi$ 
field modes, including the behaviour of the phase, in terms of the iteration
of a simple family of circle maps. The parameters of this 
family of maps are a function of time when expansion of the universe is taken into
account. With this more detailed description, we obtain a systematic study of the
efficiency of particle production as a function of the inflaton field
and coupling parameters, and we  find that for $g  \lesssim 3 \times 10^{-4}$  
the broad resonance ceases during the first stage of preheating. 
\end{abstract}

\section{Introduction}

The success of the reheating stage after inflation is crucial to most
realizations of the inflationary paradigm. 
The universe needs to
recover a temperature high enough for primordial
nuclesynthesis to take place in accordance to the usual pattern of the
standard cosmological model. Excluding scenarios designed to avoid the
extreme cooling produced by inflation (such as, for instance, the warm
inflation scenario), it is  important that the necessary
post-inflationary reheating be efficiently achieved.

The reheating mechanism was proposed as a period, immediately after inflation,
during which the inflaton field $\phi$ oscillates coherently about its ground
state and swiftly transfers its energy into ultra-relativistic matter
and radiation, here modelled by another scalar field
$\chi$. This
processes depends on the coupling between $\phi$ and $\chi$ in the
interaction Lagrangian. The classical theory of reheating was developed in
\cite{Dolgov:1982th,Abbott:1982hn,Traschen:1990sw,Dolgov:1989us}. 
The importance of broad resonance and approximations to deal
with non-perturbative effects were introduced 
in \cite{Kofman:1994rk,Boyanovsky:1996sq,Fujisaki:1995ua,Shtanov:1994ce,Berges:2002cz}.
The theory was put forward in \cite{Kofman:1997b,Greene:1997fu,Kaiser:1997hg}, 
where the analysis included the effects of expansion of the universe, see also
\cite{Yoshimura:1995gc,Fujisaki:1995dy}. This represented a shift away
from the simple picture of static Mathieu resonant bands, due to
large phase fluctuations, which behave irregularly in the non-perturbative 
regime in an expanding universe. Extensions in this setting 
include, among others, the study of metric perturbations 
and the application to string or supersymmetric
theories \cite{Henriques:2000sz,Henriques:2000xt,Bassett:1999cg,Tkachev:1998dc,Chacko:2002wr,Allahverdi:2000ss}.

The present understanding of the process (see \cite{Liddle:2000cg}) distinguishes
two parts in it. A preheating mechanism by which fluctuations of the
inflaton couple to one (or more \cite{Bassett:1998yd}) scalar fields, inducing the
resonant amplification of perturbations in the latter. 
Depending on the coupling, efficient energy transfer requires the
amplitude of the inflaton oscillation to be rather large, away from
the narrow resonance regime where only perturbations with wave numbers
in small intervals are unstable.
 As the amplitude of the perturbed field grows,
back reaction effects may have to be considered, since the frequency of
the inflaton oscillations is no longer given by its mass, but depends
also on the total number density of the perturbed field particles
through the coupling term. The first stage of preheating is the period
when these back reaction effects are negligible, and the inflaton
field dynamics is approximated by its uncoupled equations. Preheating
ends when resonant amplification terminates, either because of the
decreasing amplitude of the inflaton field or due to the back reaction
and rescattering effects of the second stage. 
After the second stage of preheating, the reheating period corresponds
to the decay of the perturbed fields as well as that of the inflaton
field that comes out of the preheating period, leaving the
universe after thermalization with the temperature required by the subsequent
processes,  namely nuclesynthesis. 

We consider a basic model describing the inflaton
field $\phi$ interacting with a scalar field $\chi$ in a flat FRW universe
\begin{eqnarray}
L = \frac{1}{2}\phi_{,i}\phi^{,i} + \frac{1}{2}\chi_{,i}\chi^{,i} - V(\phi) -  V_{int}(\phi,\chi).
\end{eqnarray}
This is the simplest model that still contains the basic features
for the understanding of particle creation in the early universe and 
one of the few models for which an analytical
study can be performed, see also
\cite{Fujisaki:1995ua,Greene:1997fu,Kaiser:1995fb}. With this in mind
we concentrate on the simplest chaotic model  with the potential
$V(\phi)=1/2 m^2_\phi \phi^2$ and interaction potential
$V_{int}(\phi,\chi) = g^2 \phi^2\chi^2$. The evolution of the
flat FRW universe is given by
\begin{eqnarray}
   \label{hubble}
  3 H^2 = \frac{8 \pi}{m_{pl}^2}\left( \frac{1}{2}\dot \phi^2 + V(\phi)+\frac{1}{2}\dot \chi^2 
  + g^2\phi^2\chi^2\right)
\end{eqnarray}
where $H=\dot R/R$ and $R$ is the FRW scalar factor.
The equations of motion  in a FRW universe for a homogeneous scalar
field $\phi$ coupled to the  $k$-mode  of the $\chi$ field 
are given by 
\begin{eqnarray}
  \label{eqphigen}
  \ddot\phi + 3H\dot\phi + \left(m_\phi^2 + g^2\chi_k^2 \right)\phi &=& 0\\
  \label{eqchigen}
  \ddot\chi_k + 3H\dot\chi_k + \omega^2_k(t)\chi_k &=& 0,
\end{eqnarray}
where $\omega_k^2(t) = k^2 /R^2 + g^2 \phi^2$ .

The rate of production of particles of a given momentum $k$ is determined by
the evolution of the perturbed field mode $\chi_k$. 
The number density $n_k(t)$ of particles with momentum $k$ can be
evaluated as the energy of that mode divided by the energy of each
particle
\begin{eqnarray}
 n_k(t) = \frac{1}{2 \omega_k(t)}\left(\vert\dot\chi_k\vert^2 + 
 \omega^2(t)\vert\chi_k\vert^2\right)-\frac{1}{2}.
\end{eqnarray}
The exponential growth $\chi_k(t)\propto\rme^{\mu_kt}$ leads to
exponential growth of the occupation numbers $n_k(t)$ and the total number density of
$\chi$-particles is given by
\begin{eqnarray}
n_\chi(t) = 
\frac{1}{4\pi^2
  R^3}\int\rmd k\, k^2 n_k(t).
\end{eqnarray}
The problem of determining the efficiency of particle production for a given model is
thus reduced to the evaluation of $\mu_k$ as a function of the
parameters of the model. 
In general this is not an easy task, and most estimates are based on numerical integrations for typical
parameter values \cite{Fujisaki:1995ua}.

However, the theory developed in \cite{Kofman:1997b} yields analytic 
results in Minkowski space time, which provide an approximate simplified 
model that works rather well in the FRW scenario. The starting point of 
the theory is the fact that, for the model (\ref {eqphigen}), (\ref {eqchigen}), 
preheating requires the amplitude of the inflaton field's oscillations, 
$\Phi (t)$, to verify $ g \Phi(t) > m_\phi$.   
This means preheating is dominated by broad resonance, and the theory
is based on the approximations that hold when $g \Phi(t)/m_\phi \gg 1$. 
The conclusions hold independently of the detailed form of the
inflaton potential away from its minimum.

In this paper we extend the formalism  of \cite{Kofman:1997b} to 
give a full description of the dynamics of the phase of the field
modes $\chi _k$, which in turn determines the evolution of the growth
factor $\mu _k$. 
In Section \ref{with.out.expansion}, we consider the case of a non
expanding universe and show that the behaviour of the phase can be
described in terms of the iteration of a simple family of circle maps.
The orbits of this family are of two possible types, depending on the
value of the perturbation amplitude, and their asymptotic behaviour,
which determines the growth factor, is always independent of the
initial condition.

In Section \ref{with.expansion} we show that the results that hold for
Minkowski space time can be used to model the phase and growth factor
evolution in a FRW universe. Expansion may be considered simply by
taking the parameters of the Minkowski equations  to be prescribed
functions of time. Using this model as an alternative to numerical
integration of the full equations, we check the estimates given
in the literature for the total number density of $\chi $ particles created
during preheating and for the duration of the first stage of
preheating as a function of the interaction parameter $g$.

\section{Phase dynamics in Minkowski space-time}
\label{with.out.expansion}

The important differences between the narrow and broad parametric resonance
mechanisms in the context of cosmological models with post-inflationary
reheating have first been noticed in the analysis of the evolution of the
amplitudes of the $\chi _k$ modes in a static universe
\cite{Boyanovsky:1996sq,Shtanov:1994ce,Biswas:2002em}.
Although the equations in Minkowski space time lack some of the fundamental
ingredients to understand the overall efficiency of the reheating process, they
can and indeed they should be considered as a toy model that sheds light on the
mechanisms at play when the expansion of the universe and other effects, such as
back reaction and rescattering, are taken into account.

This was the approach followed in Kofman {\it  et al.}  \cite{Kofman:1997b},
where an analytic theory of broad resonance in preheating was 
 established,
relying on a detailed study of broad parametric resonance driven by the 
harmonic oscillations of an inflaton field without expansion of the universe.
This study is directed towards the computation of the $k$-mode growth factors $\mu _k$,
and the phase dynamics of the $\chi _k$ modes is not explicitly derived.
However,  Kofman {\it  et al.}  do mention the basic features of this phase 
dynamics, and they use them to explain the characteristics of mode amplification
in an expanding universe, which they dubbed 'stochastic resonance' in order to
stress the difference with respect to the usual resonant bands scenario of the
Mathieu equation.

In this section we explicitly compute the phase evolution equations in
Minkowski space time and show that phase stochasticity is already present in
this model, as one of the two possible dynamical regimes, together with the
fixed phase behaviour identified in \cite{Kofman:1997b}. In this random phase
regime, in the complement of the resonant bands, the growth factor $\mu _k$ is
effectively zero, but a typical orbit undergoes random sequences of amplitude
amplifications and reductions, much like in the case of stochastic resonance.
This phenomenon is  pointed out frequently in the literature
\cite{DeMelo:2001nr,Bassett:1999mt,Finelli:2000ya} 
 and can be described analytically as one of
the consequences of the theory  when we consider the
global phase dynamics of the $\chi _k$ modes
\cite{Kofman:1997b,Fujisaki:1995ua}.

We shall start by recalling the method of  Kofman {\it  et al.} to
approximate in the broad resonance regime the solution of equation (\ref{eqchigen}) in Minkowski space-time
\begin{eqnarray}
  \label{eqchigenmst}
  \ddot\chi_k  + \omega^ 2_k(t)\chi_k=0,
\end{eqnarray}
where $\omega_k^2(t)=a_k + b\sin^2(t)$ and the time variable is now $t\to m_\phi t$.
The parameters $a_k$ and $b$ are given by  $a_k=k^2/m_\phi^2$ and
$b=g^2A^2/m_\phi^2$, where $A$ is the constant amplitude of the field $\phi $.
Typical values of the parameters are $g^2\le 10^{-6}$,
$m=10^{-6}m_{pl}$, $A=\alpha m_{pl}$, where $0 < \alpha < 1$, and thus
$b\le \alpha ^2\times 10^{6}$ \cite{Kofman:1997b,Liddle:2000cg,Linde:90}. In broad resonance, $\sqrt{b} \gg 1$.

Let the $\chi_k(t)$ be of the form 
\begin{eqnarray}
\label{wkbform}
\chi_k(t)=
\frac{\alpha_k}{\sqrt{2\omega_k(t)}}\exp\left(-\rmi\int_0^t\omega_k(s)\rmd
  s\right)+\frac{\beta_k}{\sqrt{2\omega_k(t)}}\exp\left(\rmi\int_0^t\omega_k(s)\rmd s\right),
\end{eqnarray}
where $\alpha_k$ and $\beta_k$ are constants.
Introducing (\ref{wkbform}) in to (\ref{eqchigenmst}) we obtain
\begin{eqnarray}
\label{wkbform.eq}
  \ddot\chi_k  + \omega_k^
  2(t)\left[1+\frac{1}{4}\left(\omega_k^{-1}\frac{\rmd}{\rmd t}\ln
  \omega_k\right)^2 +\frac{1}{2\omega_k}\frac{\rmd}{\rmd
  t}\left(\omega_k^{-1}\frac{\rmd}{\rmd t}\ln
  \omega_k\right) \right]\chi_k=0.
\end{eqnarray}
So
(\ref{wkbform}) approximates the solution of (\ref{eqchigenmst}) provided  that 
\begin{eqnarray}
  \left\vert \omega_k^{-1}\frac{\rmd}{\rmd t}\ln  \omega_k\right\vert\ll1,&&\\
  \left\vert \omega_k^{-1}\frac{\rmd}{\rmd t}\left(\omega_k^{-1}\frac{\rmd}{\rmd t}\ln \omega_k\right)\right\vert\ll1&&
\end{eqnarray}
hold. These are the  adiabatic conditions identified in \cite{olver:1961}.
 In the present case, $\sqrt{b} \gg 1$ and the adiabatic conditions are  fulfilled except in
 the neighbourhood  of $t_j= j \pi$, $j=0,1,\ldots$, when  $\phi(t_j)=0$
 that is, every time  the inflaton field  crosses zero.
Hence, the breakdown of the approximation given by (\ref{wkbform})
 occurs periodically, and an approximate global solution  of
 (\ref{eqchigenmst}) can be constructed from a sequence of adiabatic
 solutions 
\begin{eqnarray}
\label{chi_j}
\chi_k^j(t;\alpha_k^j,\beta_k^j)=\frac{\alpha_k^j}{\sqrt{2\omega_k(t)}}\exp\left(-\rmi\int_0^t\omega_k(s)\rmd
  s\right)+\frac{\beta_k^j}{\sqrt{2\omega_k(t)}}\exp\left(\rmi\int_0^t\omega_k(s)\rmd
  s\right),
\end{eqnarray}
where the parameters
 $(\alpha_k^j,\beta_k^j)$ for consecutive $j$ will be determined by the behaviour of he
 solution of  (\ref{eqchigenmst}) for $t$ close to $t_j$.
In a small neighbourhood of $t_j= j \pi$, $j=0,1,\ldots$   equation
(\ref{eqchigenmst}) can be approximated by 
\begin{eqnarray}
  \label{eqchigenmst.aprox}
  \ddot\chi_k  + (a_k + b (t-t_j)^2)\chi_k=0.
\end{eqnarray}
Equation (\ref{eqchigenmst.aprox}) has exact solution in the form of
parabolic cylinder functions \cite{Abramowitz:1972}. The asymptotic behaviour for large $t$
of this solutions is of the form (\ref{wkbform}), and this provides
the relation between the coefficients $(\alpha_k,\beta_k)$ of this
adiabatic approximation on either side of $t_j$. Following Kofman
{\em et al.} \cite{Kofman:1997b}, this relation can be written as
\begin{eqnarray}
\label{matrixRD}
\left[\matrix{\alpha^{j+1}_k \rme^{-\rmi\theta_k^j }\cr \beta^{j+1}_k \rme^{\rmi\theta_k^j}}\right]
=\left[\matrix{\displaystyle \frac{1}{D_{\kappa}} & \displaystyle \frac{R^*_{\kappa}}{D^*_{\kappa}}\cr \displaystyle \frac{R_{\kappa}}
{D_{\kappa}} & \displaystyle \frac{1}{D^*_{\kappa}}}\right] \left[\matrix{\alpha^{j}_k \rme^{-\rmi\theta_k^j }\cr \beta^{j}_k \rme^{\rmi\theta_k^j }}\right],
\end{eqnarray}
where $\theta_k^j=\int_0^{t_j}\omega(s)\rmd s$.
The complex numbers $R_{\kappa}$ and $D_{\kappa}$ are given by
$1/D_{\kappa} =\sqrt{1+\rho_{\kappa}^2}\exp(\rmi \varphi_{\kappa})$ and $R_{\kappa}/D_{\kappa} =-\rmi \rho_{\kappa}$
with $\rho_{\kappa}      = \exp(-\pi \kappa^2/2)$, $\kappa ^2 = a_k/\sqrt{b}$, and 
\begin{eqnarray}
  \label{varphik}
  \varphi_{\kappa} =\arg\left(\Gamma\left(\frac{1+\rmi \kappa ^2}{2}\right)\right)
  + \frac{\kappa^2}{2}\left(1+\ln\frac{2}{\kappa^2}\right).
\end{eqnarray}
The parameter $\kappa =k / \sqrt{A g m_{\phi}} \in [0,1]$ is the normalised 
wave vector of the mode, and $\sqrt{A g m_{\phi}}$ the cut-off wave vector
introduced in \cite{Kofman:1997b}. 
Since the number density of $\chi_k$ particles with momentum $k$ is
equal to $n_k=\vert\beta_k\vert^2$,  one can use
(\ref{matrixRD})  and (\ref{varphik}) to calculate the number density of particles 
$n_k^{j+1}=\vert\beta_k^{j+1}\vert^2$ after $t_j$ in terms of  $n_k^{j}=\vert\beta_k^j\vert^2$.
The growth index $\mu_\kappa^j$, defined by
$n_k^{j+1}=n_k^j\exp(2 \pi \mu_\kappa^j)$,
is given by \cite{Kofman:1997b}
\begin{eqnarray}
\label{munu}
\mu_\kappa^j = \frac{1}{2 \pi}\ln\left(1 + 2\rho_\kappa^2 -
2\rho_\kappa\sqrt{1+\rho_\kappa^2}\sin(-\varphi_\kappa +\arg \beta_k^j-\arg\alpha_k^j + 2\theta_k^j)
\right),
\end{eqnarray}
or, in terms of the phase $\nu _k^j = \arg \beta_k^j + \theta_k^j$ of the field $\chi _k$ 
when $t=t_j$,
\begin{eqnarray}
\label{munu.phase}
\mu_\kappa^j = \frac{1}{2 \pi}\ln\left(1 + 2\rho_\kappa^2 -
2\rho_\kappa\sqrt{1+\rho_\kappa^2}\sin(-\varphi_\kappa +2\nu_k^j )
\right).
\end{eqnarray}

\begin{figure}
\begin{center}
\epsfig{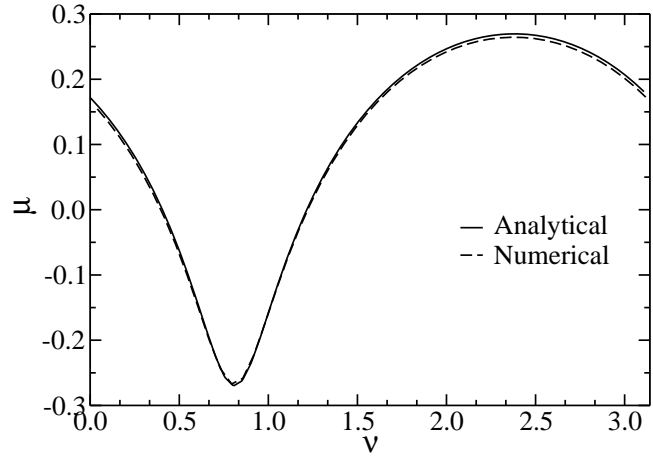}
\caption{For  $b=10^3$ and $a_k=1$ (hence $\kappa^2=10^{-3/2}$) the analytic curve for the growth factor as a function of the
  phase given by equation (\ref{munu.phase}) (solid line), and the
  numerical curve obtained from the integration of the
  equations of motion along half a period of the inflaton field
  (dashed line).
}
\label{fig1}
\end{center}
\end{figure}

In Figure
\ref{fig1} 
we show, for $b=10^3$ and $a_k=1$, the
analytic curve for the growth factor $\mu _\kappa^1$ as a function of the phase $\nu$ given by (\ref{munu.phase}) and the
numerical curve  $\mu_\kappa=\mu_\kappa(\nu)$ obtained from the numerical
integration of the full equation (\ref{eqchigenmst}) along half a period
of the inflaton field.
We see that, as $\nu$ varies in $[0, \pi]$, $\mu$ takes positive and
negative values. The phase interval for which $\mu$ is negative
depends slightly on the value of $\kappa $. 

As hinted by Kofman {\em et al.} \cite{Kofman:1997b}, equations (\ref{matrixRD})  and
(\ref{varphik}) can be used to obtain the dynamics for the phase
$\nu_k^j$. In the remaining part of this section we explicitly
compute the map $\nu_k^{j+1}=P_{b, \kappa}(\nu_k^j)$ and show that it can be approximated by a simple family of circle maps.
From (\ref{matrixRD}) one gets
\begin{eqnarray}
\label{phaseiter}
\nu_k^{j+1} = \theta (b, \kappa) + \arg \left (\sqrt{1+\rho_\kappa^2} \rme^{-\rmi\varphi_\kappa} \rme^{\rmi\nu_k^j} -\rmi\rho_\kappa\rme^{-\rmi\nu_k^j}
\right ),
\end{eqnarray}
where  $\theta (b, \kappa )=\int_0^{\pi}\omega(s)\rmd s$. An approximate expression for the  
phase map up to terms of order $\kappa ^2$ is given by
\begin{eqnarray}
\label{phaseiter_ka2}
\nu_k^{j+1} &=& P_{b, \kappa}(\nu_k^j) = 2 \sqrt{b} +
 \atan\frac{\sqrt{2} \sin \nu_k^j -\cos \nu_k^j}{\sqrt{2} \cos \nu_k^j
 - \sin \nu_k^j} \nonumber \\
 &&+ \kappa^2 /2 \left (\log {\frac{\sqrt{b}}{\kappa^2}} + 4 \log{2} + 1
\right )
- \kappa^2 \left ( \frac{ c_1 \cos {\nu_k^j}^2 + c_2 \cos \nu_k^j \sin \nu_k^j + c_3
\sin {\nu_k^j}^2}{3 - 4 \sqrt{2} \cos \nu_k^j \sin \nu_k^j }
\right ),
\end{eqnarray}
where $c_1 = 2.074 - \log 2/\kappa^2$,  $c_2 = -1.363 + 1.414 \log 2/\kappa^2$, 
and $c_3 = -0.147 - \log 2/ \kappa^2$.

\begin{figure}
\begin{center}
\epsfig{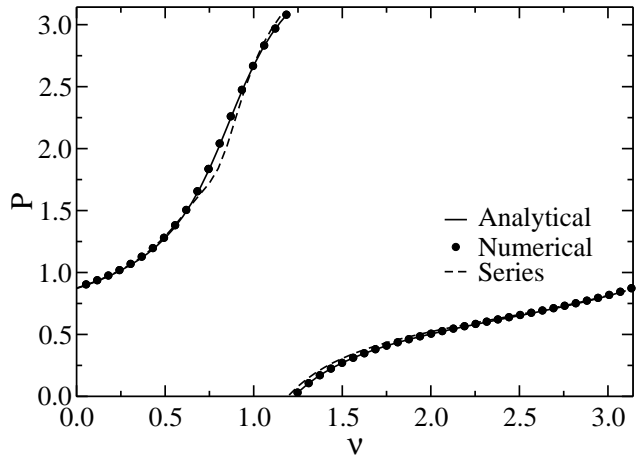}
\caption{For $b=10^3$ and $\kappa = 0.5$ the maps (\ref{phaseiter})
  (solid line) and  (\ref{phaseiter_ka2}) (dashed line). Also shown is
    the phase map obtained
  from the numerical integration of the full equations
  (\ref{eqchigenmst}) (full circle) for the same values of the parameters.}
\label{fig2}
\end{center}
\end{figure}

The maps (\ref{phaseiter}), (\ref{phaseiter_ka2}) for $b=10^3$ and $\kappa = 0.5$ are shown in
Figure \ref{fig2}, together with the phase map obtained from the numerical integration of the full
equations (\ref{eqchigenmst}) for the same values of the parameters over half a period of
the inflaton field.

The properties of the family (\ref{phaseiter}) and its approximation (\ref{phaseiter_ka2})
are best understood by looking at the behaviour of the family $P_{b,0}(\nu )$ parametrised by
$\sqrt{b}$,  
\begin{eqnarray}
\label{phaseiter_k=0}
P_{b, 0}(\nu) = 2 \sqrt{b} + \atan\frac{\sqrt{2} \sin \nu - \cos \nu}{\sqrt{2} \cos \nu - \sin \nu}
\end{eqnarray}

\begin{figure}
\begin{center}
\epsfig{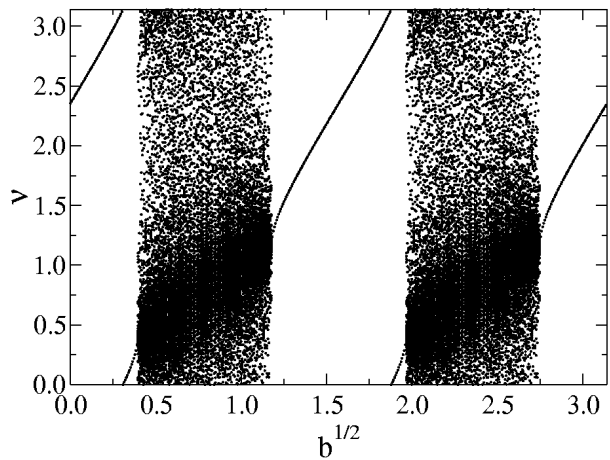}
\caption{Bifurcation diagram of the family of circle maps (\ref{phaseiter_k=0})
for  $\sqrt{b} \in [0, \pi]$.}
\label{fig3}
\end{center}
\end{figure}

This family of circle maps is periodic with period $\pi$ in the parameter $\sqrt{b}$, and its
bifurcation diagram for $\sqrt{b} \in [0, \pi]$ is shown in Figure \ref{fig3}.
We see that the map has two different
regimes. For $\tan{2 \sqrt{b}}\in [-1,1]$,
the map has a strongly attractive fixed point, and the phase
converges rapidly to this fixed point (typically in a couple of iterates).
For the remaining parameter values, the phase orbit has random
oscillations around the mean value that varies between 
$\pi/8$ and $\pi/2 -\pi/8$, or between $\pi/8 + \pi $ and $ 3\pi/2 -\pi/8$.
The fixed point equation 
\begin{eqnarray}
\label{fixedpoint_k=0}
\frac{\cos 2 \nu}{\sqrt{2}  - \sin 2 \nu} = \tan {2 \sqrt{b}} 
\end{eqnarray}
is satisfied for two values of $\nu $ for each $\sqrt{b} \in [- \pi /8, \pi /8] \cup [\pi/2 - \pi /8, \pi/2 +\pi /8]$, 
one of which corresponds to an unstable fixed point and the other to the stable fixed point shown
in the bifurcation diagram of Figure \ref{fig3}. The derivative of $P_{b,0}$ at the stable fixed point 
is 
\begin{eqnarray}
  \label{eq:dp.fp}
  P_{b, 0}'(\tilde {\nu })=\frac{1}{3-2\sqrt{2}\sin 2\tilde {\nu }}\le 1,
\end{eqnarray}
where the stable fixed point $\tilde {\nu } \in
[-\pi/2 -\pi/8,\pi/8]\cup [\pi/2-\pi/8,\pi + \pi/8]$. The equality $ P_{b, 0}'(\tilde {\nu })=1$
is obtained at the boundaries of the random phase region, but the derivative
decreases rather sharply into the stable fixed point region, where  $
P_{b, 0}'(\bar {\nu })< 0.5$ for most values.

\begin{figure}
\begin{center}
\epsfig{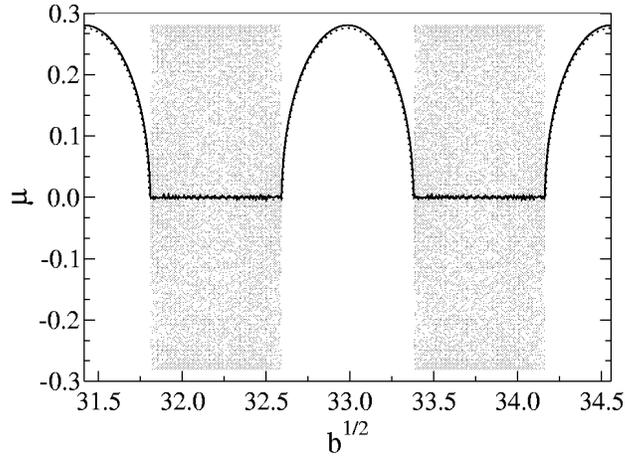}
\caption{The asymptotic value of $\mu _0$ as a function of $b$ for
  $\sqrt{b}\in [10 \pi,11\pi]$ computed 
analytically from equations (\ref{phaseiter_k=0}), (\ref{munu.phase})
(full line) and numerically  from the integration of the
full equations (\ref{eqchigenmst}) (dotted line).
We have taken  $j_{max}=200$, and the two lines almost overlap. Also shown (in grey)
are all the values of $\mu_0^j$, $j=100,101,\ldots,200$.}
\label{fig4}
\end{center}
\end{figure}

The asymptotic value of growth factor $\mu_0$ for the $k=0$ mode as a function of $b$ can then be computed from 
equation (\ref{munu.phase}) evaluated at the fixed point or averaged
over the random orbits, for values of $\sqrt{b}\in [- \pi /8, \pi
/8]\cup [\pi/2 - \pi /8, \pi/2 +\pi /8]$
 or in the complement of this interval, respectively. 
In Figure \ref{fig4} we show a plot of the asymptotic value of $\mu _0$ as a function of $b$ computed 
analytically  from equations (\ref{phaseiter_k=0}), (\ref{munu.phase}),
as described, and numerically from the integration of the full equations 
(\ref{eqchigenmst}). We see that the stability regions identified in 
\cite{Kofman:1997b} correspond to the random wandering of the phase $\nu $ of the field $\chi $
at consecutive $t_j = j \pi$ around its average values. The random phase regimes correspond to different probability measures in the phase interval, for all of which the average value of $\mu $ is zero.

\begin{figure}
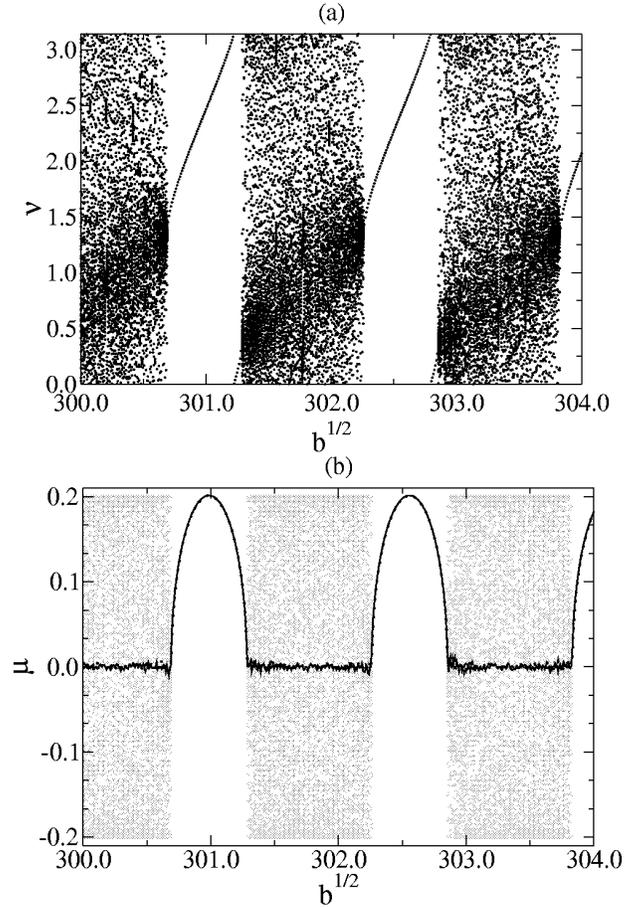

\begin{center}
\epsfig{file=figure5a.eps, height=6cm}
\epsfig{file=figure5b.eps, height=6cm}
\caption{(a) Bifurcation diagram for the map $P_{b,\kappa}$ with
  $\sqrt{b}\in[300,304]$ and $\kappa = 1/2$. (b) The asymptotic value
  of $\mu_\kappa$ as a function of $\sqrt{b}$ for the same values of
  the parameters  computed 
analytically  from equations (\ref{phaseiter}), (\ref{munu.phase})
  (full line), and numerically from the integration of the full equations 
(\ref{eqchigenmst}) (dotted line). We have taken $j_{max}=200$, and the two
  lines almost overlap. Also shown (in grey) are all the values of
  $\mu_\kappa^j$ , $j=100,101,\ldots,200$.}
\label{fig5}
\end{center}
\end{figure}

For other values of $\kappa $, the global dynamics shares the qualitative properties of the family
$P_{b,0}$. In Figure \ref{fig5} we show the same information of Figures \ref{fig3} and \ref{fig4} obtained for 
$P_{b,\kappa}$ with $300\le \sqrt{b}\le 304$ and $\kappa = 1/2$.

\begin{figure}
\begin{center}
\epsfig{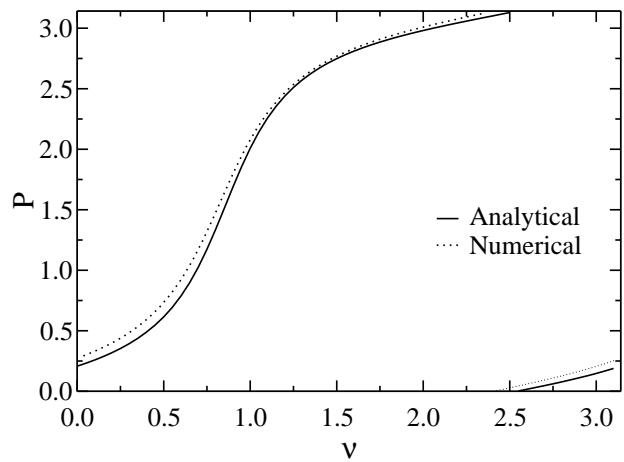}
\caption{The approximate phase
map(solid line) and the phase map computed numerically from equations
(\ref{eqchigenmst}) (dotted line) for $b=10$ and $\kappa =1/2$.}
\label{fig6}
\end{center}
\end{figure}

Equation (\ref{phaseiter}) provides a good approximation to the
exact phase dynamics even for moderate values of $b$. In Figure \ref{fig6} we show the approximate phase map
(\ref{phaseiter}) and the phase map computed numerically from the
full equation (\ref{eqchigenmst}) for $b=10$ and $\kappa =1/2$.
This, together with the rapid relaxation rates of the phase dynamics
in the fixed point regime, is the reason why equations (\ref{munu.phase}), (\ref{phaseiter}) are still useful to obtain estimates for the growth number in FRW space-time. The existence of two simple regimes for the phase dynamics, one of them characterised by rapid relaxation to a fixed point and the other by random wandering of the phase with a well defined probability density, shows that the system keeps no record of the initial phase and, in this sense, has no memory. This explains why the evolution of the occupation number is independent of the initial phase, while the growth factor per period $\mu_{\kappa}^j$ depends strongly on the phase $\nu_k^j$.
We shall come back to this point later.

\section{Dynamics in an expanding universe}
\label{with.expansion}

In this section we will show that equations (\ref{munu.phase}),
(\ref{phaseiter})  can be used to compute the growth factor of the
$\chi$ field modes during the first stage of the preheating period in a flat FRW.

It is well known \cite{Kofman:1997b}, that the coherence of the
oscillations of the $\phi$ field is not disturbed until the energy
density of the $\chi$ field significantly contributes to
(\ref{hubble}), (\ref{eqphigen}) and back-reaction and rescattering effects start to
change the mechanism of the growth of the occupation number of the
produced particles. More precisely,  Kofman {\it  et al.} show in
\cite{Kofman:1997b} that these effects are negligible during the first
preheating stage, that ends when the total number density $n_\chi$ of 
$\chi $ particles satisfies
\begin{eqnarray}
\label{cutoffcond}
n_\chi (t) \approx \frac{m_\phi^2 \Phi(t)}{g},
\end{eqnarray}
where $\Phi(t)$ is the varying amplitude of the inflaton field $\phi$. 

In the first stage of preheating, equations (\ref{hubble}), (\ref{eqphigen}) decouple from (\ref{eqchigen}), and the evolution of the inflaton field and of the scale factor $R(t)$ is given in good approximation by \cite{Kofman:1997b}
\begin{eqnarray}
  \label{FRWphiR}
  \phi (t)= \Phi(t) \sin t,\qquad \Phi(t)=\frac{m_{pl}}{3 (\pi /2 + t)},\qquad 
R(t)= \left (\frac{2 t}{\pi}\right )^{2/3}.
\end{eqnarray}

Equation (\ref{eqchigen}) can be reduced to the form 
(\ref{eqchigenmst}) through the change of variable 
$X_k = R^{3/2}\chi_k$, yielding
\begin{eqnarray}
\label{eqX}
\ddot X_k + \varpi_k^2(t)X_k=0,
\end{eqnarray}
where $\displaystyle \varpi^2 =  \frac{k^2}{m_\phi^2R(t)^2} + \frac{g^2\phi(t)^2}{m_\phi^2}
+ \frac{1}{m_\phi^2}\delta$ and the last term is very small after inflation and will be disregarded. 

We see that the $\chi _k$ modes are still governed by an equation of the form  (\ref{eqchigenmst}) like the one we considered in Minkowski space time, but now the parameters $a_k$ and $b$ change with time. By definition, the preheating period ends when $g \Phi(t)/m_\phi \simeq 1$, and so, during preheating, the rate of variation of those parameters and the oscillations of the inflaton field are much slower than the oscillations of the $\chi _k$ modes.  As pointed out in \cite{Kofman:1997b}, the basic assumptions for the approximation developed for Minkowski space time are thus still valid in preheating, 
and the changes in occupation numbers $n_k$ will occur at $t=j \pi$ with exponential growth rate given by  (\ref{munu.phase}),  
provided that the decreasing amplitude of the perturbations and the redshift of the wave numbers are taken into account. 
Hence, Kofman {\it  et al.} model particle production in the first stage of preheating through
\begin{eqnarray}
\label{mu.exp}
\mu_{\kappa_j}^j = \frac{1}{2 \pi}\ln\left(1 + 2\rho_{\kappa_j}^2 -
2\rho_{\kappa_j}\sqrt{1+\rho_{\kappa_j}^2}\sin(-\varphi_{\kappa_j} +2\nu_\kappa^j )
\right),
\end{eqnarray}
where $\kappa_j=k/(R(t_j) \sqrt{g m_\phi \Phi(t_j)})$.

We may also think of the phase dynamics as being essentially governed by the iteration of equations  (\ref {phaseiter}), but now the parameter $b$ 
decreases in time, crossing the strips associated to the random phase and fixed point regimes, and giving rise to non trivial phase dynamics. 
Kofman {\it  et al.}  describe the effects of the variation of the inflaton
amplitude as implying random phase dynamics, and build their estimates for
the total number density evolution $n_\chi (t)$ from (\ref {mu.exp}), treating the phase $\nu_\kappa^j$ as a random variable.

We shall extend the approach of \cite{Kofman:1997b} to study the phase dynamics in an expanding universe, taking the phase iteration map to be given by
\begin{eqnarray}
\label{phaseiter.exp}
\nu_{k}^{j+1} = \theta (b_j, \kappa_j) + \arg \left (\sqrt{1+\rho_{\kappa_j}^2} \rme^{-\rmi\varphi_{\kappa}} \rme^{\rmi\nu_k^j} -\rmi\rho_{\kappa_j}\rme^{-\rmi\nu_k^j}
\right ),
\end{eqnarray}
where $\sqrt{b_j} = g \Phi (t_j)/m_\phi$.

\begin{figure}
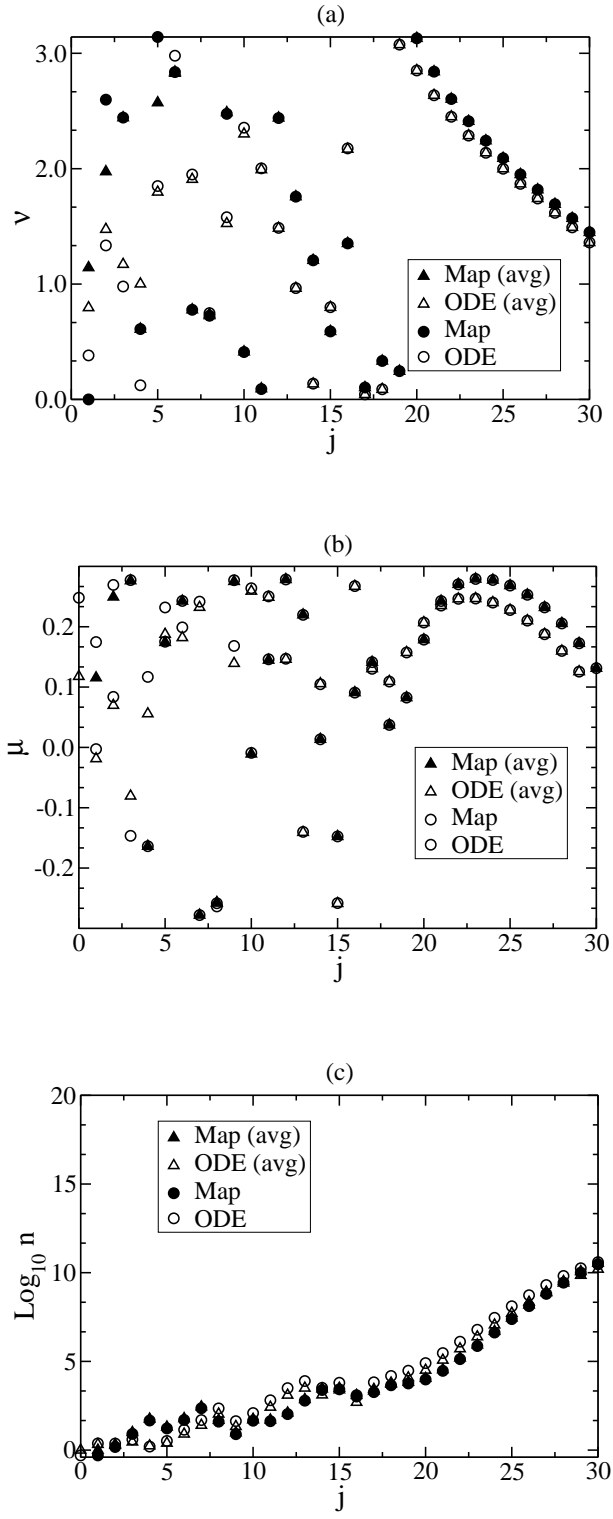

\begin{center}
\epsfig{file=figure7a.eps, height=6cm}
\vspace{1cm}
\\

\epsfig{file=figure7b.eps, height=6cm}
\vspace{1cm}
\\

\epsfig{file=figure7c.eps, height=6cm}
\caption{
For $b_0= 5 \times 10^3$ and $\kappa_0=0.1$, the phase (a),
growth factor (b) and occupation number (c) with initial conditions
corresponding to $n_k^0 = 1/2$ and $\nu _k^0 = 0$. The values
obtained from the iteration of equations  (\ref {phaseiter.exp}),
(\ref {mu.exp}) are plotted as full circles, and the values given by
numerical integration of equations (\ref {eqX}) for the same initial
conditions and parameter values are plotted as open circles. Iteration
and integration were carried out until the end of preheating when
$\sqrt{b(t_j)} \approx 1$. Also shown are the values of these same
quantities averaged over the initial phase $\nu _k^0$ (full triangles
for the iterated maps (\ref {phaseiter.exp}), (\ref {mu.exp}) and open
triangles for the numerical values).}

\label{fig7}
\end{center}
\end{figure}

Equations (\ref {phaseiter.exp}), (\ref {mu.exp}), provide an
alternative to numerical integrations of the full equations do compute
the occupation number of a given mode as a function of time. The phase,
growth factor and occupation number for a typical orbit (we have taken
$b_0=5 \times 10^3$ and $\kappa_0=0.1$) as obtained from the iteration
of equations  (\ref {phaseiter.exp}), (\ref {mu.exp}) are shown in
Figure \ref{fig7}, where the values given by numerical integration of
equations (\ref {eqX}) for the same initial conditions and parameter
values are also plotted. Iteration and integration were carried out
until preheating ends with $\sqrt{b(t_j)} \approx 1$. We see
'reminiscences' of the phase dynamics of the Minkowski model. In
particular, the fixed point regime interval is
clearly visible after the first few $\phi $ oscillations. Also shown
are the values of these same quantities averaged over the initial
phase $\nu _k^0$. We see that, due to the characteristics of the phase
dynamics, the efficiency of particle production is insensitive 
to the initial phase of the field mode $\chi _k$, in spite of the
strong dependence of the growth factor per period on the phase $\nu
_k^j$.

\begin{figure}
\begin{center}
\epsfig{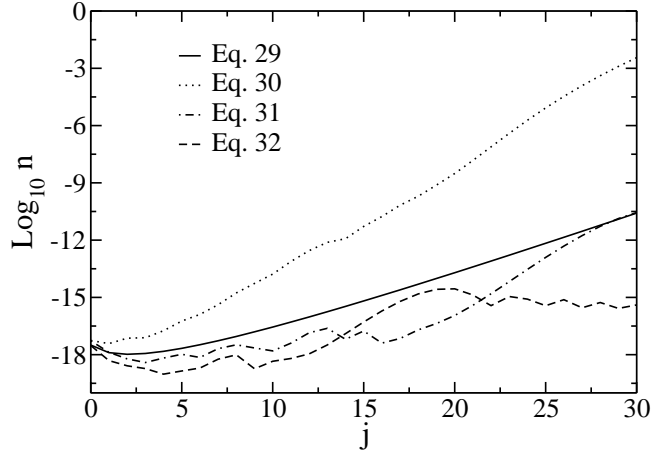}
\caption{
For $b_0= 5\times 10^{3}$  (hence $g= 0.3 \times 10^{-3}$) the curves corresponding to equations  (\ref{nchirussos}) (full line), (\ref{nchi1}) (dotted line), (\ref{nchi2}) (slash-and-dot line) and (\ref{nchirussos0}) (slashed line).
The evolution was computed until the end of preheating when $\sqrt{b_{30}}
\simeq 1$.}
\label{fig8}
\end{center}
\end{figure}

We shall now use equations  (\ref {phaseiter.exp}), (\ref {mu.exp}) above to look at the total number density $n_\chi (t)$ and check 
the estimates given in \cite{Kofman:1997b}.
We consider the contributions to $n_\chi (t)$ of every mode such that  $k \in [0, k_*(0)=\sqrt{g \Phi(0) m_\phi}]$ at $t=0$, 
and compute for each $j$ the leading mode's growth factor, $\mu_l^j$, given by

\begin{eqnarray}
\label{leading}
\mu_l^{j} = max_{\kappa ^2 \in [0, 1]} \left\{\frac{1}{2 \pi} \sum _{\nu^0 \in [0, 2 \pi]}  \mu ^j_{\kappa _j} 
( P^{(j)}_{b_j, \kappa _j}(\nu ^0))\right\},
\end{eqnarray}
with  $b_j =5 \times 10^3 (\Phi (t_j)/\Phi (0))^2$, $\kappa_j ^2=\kappa ^2\Phi
(0)/\left(\Phi (t_j) R^2(t_j)\right)$, and $P^{(j)}_{b_j, \kappa _j}$ the $j$-th
iterate of the map (\ref{phaseiter.exp}), where the parameters
$\kappa $ and $b$ must be updated at each iteration. 
Then, whereas the estimate in \cite{Kofman:1997b} is
\begin{eqnarray}
\label{nchirussos}
n_\chi (t) = \frac{(g m_\phi \Phi(0))^{3/2}}
{64 \pi^2 R^3(t) \sqrt{0.13 \pi t}} \exp(2\times 0.13 t),
\end{eqnarray}
we have
\begin{eqnarray}
\label{nchi1}
n_{\chi} (t_j)= \frac{(g m_\phi \Phi(0))^{3/2}}
{64 \pi^2 R^3(t) \sqrt{ \pi ^2\sum_{i=1}^j \mu_l^i}} \exp\left(2 \pi\sum _{i=1}^j \mu_l^i\right),
\end{eqnarray}
or 
\begin{eqnarray}
\label{nchi2}
n_{\chi} (t_j)= \frac{(g m_\phi \Phi(0))^{3/2}}
{64 \pi^2 R^3(t) \sqrt{ \pi ^2\sum_{i=1}^j \mu_*^i}} \exp\left(2 \pi\sum _{i=1}^j \mu_*^i\right),
\end{eqnarray}
if instead of actually determining the leading mode we assume that it corresponds to
$k_*/2$.
The curves corresponding to equations (\ref {nchirussos}), (\ref {nchi1}), (\ref {nchi2})
are shown in Figure \ref{fig8}.
We can still improve on the estimates given by (\ref {nchi1}) or (\ref {nchi2}). Since
\begin{eqnarray}
\label{nchirussos0}
n_\chi (t) = \frac{(g m_\phi\Phi(t))^{3/2}}{64 \pi ^2 R^3(t)} \int _0^{1}
\kappa^2 n_\kappa(t) \rmd \kappa,
\end{eqnarray}
we may use (\ref {phaseiter.exp}), (\ref {mu.exp}) to compute this
integral numerically from $n_k (t_j) = 1/2 \exp \left(2 \pi \sum_{i=1}^j \mu _{\kappa_i}^i\right)$.
The result is also shown in Figure \ref{fig8}, and we see that for this value 
of $g$ the estimates of \cite{Kofman:1997b} are very accurate.

\begin{figure}
\begin{center}
\epsfig{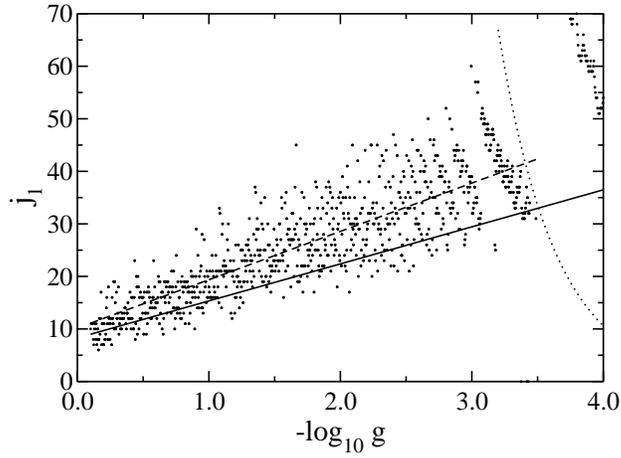}
\caption{Duration of the first 
stage of preheating $j_1$ as a function of $g$ for $10^{-4}\le g\le
10^{-1}$ (dots). The linear least square
approximation (slashed line), are also shown the estimate given in
\cite{Kofman:1997b} (full line) and the cut-off curve
defined by $b_j=1$ (dotted line). The values of $j_1$ that lie above
the cut-off curve $b_j=1$ were not considered in the least square approximation}.
\label{fig9}
\end{center}
\end{figure}

Finally, we have used again (\ref {phaseiter.exp}), (\ref {mu.exp}) 
together with (\ref {nchirussos0}) to obtain a systematic study both of the behaviour of
$n_\chi (t)$ as a function of the parameter $g$, and  of the duration of the first 
stage of preheating as defined by (\ref {cutoffcond}), as a function of $g$. The 
results for $g$ in the range $10^{-4}\le g\le 10^{-1}$ are shown in
Figure \ref{fig9}, where we have also plotted the result of the estimates of
\cite{Kofman:1997b} for these quantities.  
We see that the duration of the first stage of preheating depends rather
irregularly on the parameter $g$, and that the estimates of \cite{Kofman:1997b}
 (full line in the figure) provide a good lower bound for most values of $g$.
 The least squares linear fit (slashed line) yields a slightly larger 
value for the typical duration of the first stage. 
On the other hand, the value of $g$ below which preheating always ceases 
during the first stage was found to be $g = 3\times 10^{-4}$, as predicted in
\cite{Kofman:1997b}.

\section{Conclusion}

In the simplest preheating scenario, where the coherent oscillations of the uncoupled 
inflaton field drive the amplification of the mode amplitudes of a field
$\chi $, we consider first the broad resonance regime in Minkowski space-time and
use the theory of scattering in parabolic potentials developed in
\cite{Kofman:1997b}  to obtain the map whose iteration governs the
phase dynamics of the modes $\chi _k$. It is well known that the phase dynamics,
the consecutive values of the phase of the $\chi _k$ fields at the times $t_j$
when the inflaton amplitude goes through zero, determine the growth rates of the
modes. In this work we show that the features of this phase dynamics are given by the
properties of a simple family of circle maps. The orbits of this family of maps
are of two types, rapid convergence to fixed point solutions, and random
oscillations around an average value. Hence, the 'stochastic resonance'
identified in \cite{Kofman:1997b}  in the dynamics of an expanding universe 
is also present in the absence of expansion. The fixed phase and stochastic regimes
occur  in consecutive intervals of the value of the forcing amplitude. 
In the first case, the fixed point is always associated with a positive
value of the growth factor $\mu _j = 1/(2\pi)\log\left(n^{j+1}/n^{j}\right) $ that 
controls the growth of the number of particles $n^{j}$ in each mode for $t=t_j$.
Thus, in this case, the average growth of the occupation numbers of the modes 
is exponential. In the second case, we show that the phase sampling is always such 
that the average growth factor is zero.

We then consider the case of an expanding universe, with the assumptions that
hold in the first stage of preheating, and show  that the equations
for the phase dynamics and the growth number derived for Minkowski space time
still provide a good approximation of the true solutions, once the decay
of the inflaton amplitude is taken into account. Moreover, the qualitative behaviour
of the phase and growth number evolution is reminiscent of the behaviour found
in the case without expansion, in the sense that it can be interpreted as 
a random phase regime followed by a slowly varying phase regime where occupation
number growth is approximately exponential. These two regimes occur as the inflaton 
decay slows down and the perturbation amplitude crosses more and more slowly the intervals 
that give rise to fixed phase behaviour.
  
We use this approximation to obtain a systematic study of the behaviour of the
total number density of created particles over time, and of the end of the
first stage of preheating as a function of the $\phi-\chi$ coupling parameter $g$.
Comparison with the estimates presented in \cite{Kofman:1997b} show an overall
good agreement.

\section{Acknowledgements}

The authors are grateful to David Wands for helpful discussions.
We also acknowledge the financial support of Funda\c c\~ao para a Ci\^encia
e a Tecnologia under the grant number POCTI/FNU/49511/2002.


\newpage

\end{document}